\magnification 1050
\baselineskip 14 pt

\def \P {{\bf P}}
\def \E {{\bf E}}
\def \F {{\bf F}}
\def \c {{\bf c}}
\def \D {{\bf D}}
\def \bDo {\overline {\cal R}}
\def \L {{\cal L}}
\def \R {{\bf R}}
\def \Do {\R \times \R^+}
\def \up {\Upsilon}

\def \lx {\lambda_x}
\def \ly {\lambda_y}

\def \la {\lambda}
\def \al {\alpha}

\def \noi {\noindent}
\def \bU {\overline U}
\def \bD {\overline D^*}
\def \bk {\overline k}
\def \bt {\overline \theta}
\def \bc {\overline c}

\def \bdel {\overline \delta}
\def \btau {\overline \tau}

\def \sect#1{\bigskip  \noindent{\bf #1} \medskip }
\def \subsect#1{\bigskip \noindent{\it #1} \medskip}
\def \thm#1#2{\medskip \noindent {\bf Theorem #1.}   \it #2 \rm \medskip}
\def \prop#1#2{\medskip \noindent {\bf Proposition #1.}   \it #2 \rm \medskip}
\def \cor#1#2{\medskip \noindent {\bf Corollary #1.}   \it #2 \rm \medskip}
\def \pf {\noindent  {\it Proof}.\quad }
\def \lem#1#2{\medskip \noindent {\bf Lemma #1.}   \it #2 \rm \medskip}

\def \rem#1{\medskip \noindent {\bf Remark #1.}}

\def\sqr#1#2{{\vcenter{\vbox{\hrule height.#2pt\hbox{\vrule width.#2pt height#1pt \kern#1pt\vrule width.#2pt}\hrule height.#2pt}}}}

\def \square{\hfill\mathchoice\sqr56\sqr56\sqr{4.1}5\sqr{3.5}5}

\def \qed {$\square$ \medskip}

\nopagenumbers

\headline={\ifnum\pageno=1 \hfill \else \hfill {\rm \folio} \fi}

\centerline{\bf Life Insurance Purchasing to Maximize Utility of Household Consumption}
\bigskip

\centerline{Erhan Bayraktar}

\centerline{Virginia R. Young} \bigskip

\centerline{Department of Mathematics, University of Michigan}
\centerline{Ann Arbor, Michigan, 48109} \bigskip

\centerline{Version:  11 January 2013} \bigskip

\noindent{\bf Abstract:}  We determine the optimal amount of life insurance for a household of two wage earners.  We consider the simple case of exponential utility, thereby removing wealth as a factor in buying life insurance, while retaining the relationship among life insurance, income, and the probability of dying and thus losing that income. For insurance purchased via a single premium or premium payable continuously, we explicitly determine the optimal death benefit.  We show that if the premium is determined to target a specific probability of loss per policy, then the rates of consumption are {\it identical} under single premium or continuously payable premium.  Thus, not only is equivalence of consumption achieved for the households under the two premium schemes, it is also obtained for the insurance company in the sense of equivalence of loss probabilities.

\bigskip

\noindent{\it Keywords:} Life insurance, utility maximization, optimal consumption, optimal investment, exponential utility.

\sect{1. Introduction}

The problem of finding the optimal strategy for purchasing life insurance has not been studied nearly as much as the corresponding problem for life annuities; see Milevsky and Young (2007) and the many references therein for research dealing with life annuities.  Notable exceptions are the seminal papers of Richard (1975) and Campbell (1980) and the more recent papers of Pliska and Ye (2007), Huang and Milevsky (2008), Kraft and Steffensen (2008), Nielsen and Steffensen (2008), Wang et al.\ (2010), Kwak, Shin, and Choi (2011), Bruhn and Steffensen (2011), and Egami and Iwaki (2011).  We highlight the last two articles because that work is closely related to the work in this paper.  Bruhn and Steffensen (2011) find the optimal insurance purchasing strategy to maximize the utility of consumption for a household, in which utility of consumption is a {\it power} function. Thus, the optimal death benefit is related to the wealth of the household.  Similarly, Egami and Iwaki (2011) maximize household utility of consumption and terminal wealth at a fixed time and find that the optimal death benefit is a function of wealth because of the nature of their utility function.

In this paper, we focus on the idea that, when maximizing the utility of household consumption, life insurance serves as {\it income replacement}; therefore, the amount purchased should be independent of the household's wealth.  Indeed, when a wage earner dies, the wealth of the household remains, but the income of that wage earner is lost; life insurance can replace that lost income.  Granted, life insurance is often awarded to the beneficiary in a single payment, but the beneficiary can use that payment to buy an annuity, which more clearly replaces lost income.

To obtain an optimal amount of life insurance that is independent of the wealth of the household, we assume that utility of consumption is {\it exponential}. The resulting optimal death benefit is a function of the rates of mortality and rates of income of the household but not of its wealth.  An added bonus of using exponential utility is that the optimization problem greatly simplifies, and we explicitly express the optimal death benefit, the optimal rates of consumption (both before and after the death of a wage earner), and the optimal investment in the risky asset.

The rest of the paper is organized as follows: In Section 2, we model the household as two wage earners; the household wishes to maximize its (exponential) utility of consumption until the second death, with the possibility of purchasing life insurance that pays at the first death.  Additionally, we allow the household to invest in a risky and a riskless asset.  In Sections 3 and 4, we solve the optimization problem in the case of premium paid with a single premium or payable continuously until the first death, respectively. We find {\it explicit} expressions for the optimal amount of death benefit for the household, and we determine how that death benefit varies with the underlying parameters.  In Section 5, we compare the optimal rates of consumption from Sections 3 and 4 and show that they are {\it identical} if premium is determined by targeting a specific probability of loss per policy. 

For those readers who wish to skip the mathematical content of the paper, Section 6 provides a numerical example to illustrate our results, and Section 7 concludes the paper with a detailed summary of those results. We also have available an Excel spreadsheet at

http://www.math.lsa.umich.edu/$\sim$erhan/OptimalLifeInsurance.xlsx

\noi that the interested reader can use to calculate the optimal amount of life insurance for his or her household.

\sect{2.  Financial Background, Statement of Problem, Verification Lemma}

In this section, we present the financial and insurance market for the household.  Then, we state the optimization problem that this household faces.  Finally, we present a verification lemma that we use to solve the optimization problem.

\subsect{2.1. Financial market}

The household invests in a Black-Scholes financial market with one riskless asset earning at the rate $r \ge 0$ and one risky asset whose price process $S = \{ S_t \}_{t \ge 0}$ follows geometric Brownian motion:
$$
dS_t = \mu \, S_t \, dt + \sigma \, S_t \, dB_t,
$$
in which $B = \{ B_t \}_{t \ge 0}$ is a standard Brownian motion on a filtered probability space $(\Omega, {\cal F}, \F = \{ {\cal F}_t \}_{t \ge 0}, \P)$, with $\mu > r$ and $\sigma > 0$.  (In courses on the theory of interest, actuaries call $r$ the {\it force of interest}.)

Let $W_t$ denote the wealth of the household at time $t \ge 0$.  Let $\pi_t$ denote the dollar amount invested in the risky asset at time $t \ge 0$.  Finally, let $c_t$ denote the continuous rate of consumption of the household at time $t \ge 0$.

A consumption policy $\c = \{ c_t \}_{t \ge 0}$ is admissible if it is an $\F$-progressively measurable process satisfying $\int_0^t | c_s | \, ds < \infty$ almost surely, for all $t \ge 0$.  An investment policy $\Pi = \{ \pi_t \}_{t \ge 0}$ is admissible if it is an $\F$-progressively measurable process satisfying $\int_0^t \pi^2_s \, ds < \infty$ almost surely, for all $t \ge 0$.  Furthermore, we require that the policies are such that $W_t > - \infty$ for all $t \ge 0$ with probability 1. For example, this restriction prohibits the household from borrowing an infinite amount of wealth and consuming it all instantly.

The household receives income from two sources:  While family member $(x)$ is alive, the household receives income at the continuous rate of  $I_x \ge 0$.  Similarly, while family member $(y)$ is alive, the household receives income at the continuous rate of $I_y \ge 0$.  Denote the future lifetime random variables of $(x)$ and $(y)$ by $\tau_x$ and $\tau_y$, respectively.  We assume that  $\tau_x$ and $\tau_y$ follow independent exponential distributions with means $1/\lx$ and $1/\ly$, respectively.  (In other words, $(x)$ and $(y)$ are subject to constant forces of mortality, $\lx$ and $\ly$, respectively.)  Let $\tau_1$ denote the time of the first death of $(x)$ and $(y)$, that is, $\tau_1 = \min( \tau_x, \tau_y)$.  Let $\tau_2$ denote the time of the second death, that is, $\tau_2 = \max( \tau_x, \tau_y)$.

The household buys life insurance that pays at time $\tau_1$, the moment of the first death.  This insurance acts as a replacement for the income lost when the wage earner dies.  In this time-homogeneous model, we assume that a dollar death benefit payable at time $\tau_1$ costs $H$ at any time when both $(x)$ and $(y)$ are alive.  Write the single premium as follows:
$$
H = (1 + \theta) \bar A_{xy} = (1 + \theta) {\lx + \ly \over \lx + \ly + r},
\eqno(2.1)
$$
in which $\theta \ge 0$ is the proportional risk loading.  Assume that $\theta$ is small enough so that $H < 1$ because if $H \ge 1$, then the buyer would be foolish to pay a dollar or more for each dollar of death benefit. The inequality $H < 1$ is equivalent to $r > \theta (\lx + \ly)$.

In this section and in Section 3, we suppose that the premium is payable at the moment of the contract; as stated above, $H$ is the single premium.  In Section 4, we consider the case for which the insurance premium is payable continuously at the rate $h = (1 + \overline \theta) {\bar A_{xy} \over \bar a_{xy}} = (1 + \bt) (\lx + \ly)$ until time $\tau_1$.

Let $D_t$ denote the amount of death benefit payable at time $\tau_1$  purchased at or before time $t \ge 0$.  A life insurance purchasing strategy $\D = \{ D_t \}_{t \ge 0}$ is admissible if it is an $\F$-progressively measurable, non-decreasing process with $W_t > -\infty$ for all $t \ge 0$ with probability 1.  Thus, with single-premium life insurance, wealth follows the dynamics
$$
\left\{
\eqalign{
dW_t &= (r W_{t-} + (\mu - r) \pi_{t-} + I_x + I_y  - c_{t-}) \, dt + \sigma \, \pi_{t-} \, dB_t - H \, dD_t, \quad 0 \le t < \tau_1, \cr
W_{\tau_1} &= W_{\tau_1-} + D_{\tau_1-} \, , \cr
dW_t &= (r W_t + (\mu - r) \pi_t + I_x \, {\bf 1}_{\{\tau_1 = \tau_y \}}  + I_y \, {\bf 1}_{\{\tau_1 = \tau_x \}}  - c_t) \, dt + \sigma \, \pi_t \, dB_t, \quad \tau_1 < t < \tau_2.
}
\right.
\eqno(2.2)
$$
A few words of explanation of the expressions in (2.2) are in order.  Because we allow the life insurance purchasing strategy to jump due to lump-sum purchases, we write the subscript $t-$ instead of $t$ to denote the values of  the corresponding process {\it before} any such jump.  Before the first death, the household's wealth increases at the rate of interest earned on the riskless asset, $r(W_{t-} - \pi_{t-})$, the drift on the risky asset, $\mu \, \pi_{t-}$ (modified by the random component $\sigma \, \pi_{t-} \, dB_t$), and the rates of income, $I_x$ and $I_y$. Also, before the first death, the household's wealth decreases with consumption and any purchase of life insurance.

At the moment of the first death, the wealth of the household immediately increases by the death benefit $D_{\tau_1-}$. After that death, the wealth process is much as it was before, except for the loss of the income of the family member who died.  Also, the optimal amount of wealth invested in the risky asset and the rate of consumption chosen by the remaining family member may change.

\subsect{2.2. Utility of consumption}

We assume that the household seeks to maximize its expected discounted utility of consumption between now and $\tau_2$, the time of the second death, by optimizing over admissible controls $(\c, \Pi, \D)$.  The corresponding value function is given by
$$
U(w, D) = \sup_{(\c, \Pi, \D)} \E^{w, D} \left[ \int_0^{\tau_2} e^{- r t} \, u(c_t) \, dt \, \bigg| \, \tau_1 > 0 \right],
\eqno(2.3)
$$
in which $\E^{w, D}$ denotes conditional expectation given that $W_{0-} = w$ and $D_{0-} = D$.  Here, $r$ is the discount factor for the household, which we assume equals the rate of return for the riskless asset. The function $u$ is the utility of consumption; we assume throughout this paper that
$$
u(c) = - \, {1 \over \al} \, e^{- \al c},
\eqno(2.4)
$$ for a constant $\al > 0$, which is called the {\it coefficient of absolute risk aversion}. Thus, exponential utility is equivalent to constant absolute risk aversion.

To make this paper somewhat complete, we now present a way a household might determine its constant absolute risk aversion $\al$.  It is a method we borrow from Bowers et al.\ (1997). Suppose a household faces the possible loss of \$10,000 with probability 0.01 and no loss with probability 0.99.  In lieu of that random loss, the household is offered insurance for a fixed premium, and suppose $P$ is the maximum insurance premium that the household would pay in exchange for the possibility of losing \$10,000.  That is, the household is indifferent between (i) paying $P$ and receiving full reimburse for the loss and (ii) buying no insurance for the loss and paying for it out of pocket.  Then, by expected utility theory, if the household's current wealth is $w$, the expected utility of these two alternatives are equal:
$$
e^{-\al(w - P)} = 0.01 e^{-\al(w - 10000)} + 0.99 e^{-\al w},
$$
or equivalently
$$
P =  {1 \over \al} \ln \left( 0.01 e^{10000 \al} + 0.99 \right) .
$$
For $P \in (100, \, 10$,$000)$, this equation has a unique solution $\al > 0$; see Gerber (1974) for more on the exponential premium principle.

\rem{2.1} One can modify our set up to model a household composed of one income earner and other members. Indeed, let $I_x > 0$ and $I_y = 0$.  In that case, it would be reasonable to suppose that the household will buy insurance that pays only at time $\tau_x$, instead of $\tau_1$.

Additionally, one can modify our set up to model a family composed of two income earners and several children. In that case, let $I_x > 0$ and $I_y > 0$. One could consider a variety of life insurance products in this case, for example, separate policies on $(x)$ and $(y)$. Also, it would make sense to maximize the utility of consumption until the third ``death,'' namely, when the two wage earners and all the children have died.

We leave the details of these adaptations to the interested reader and proceed with the case of a two-party household of two earners. We believe that the qualitative results we find here will hold in the cases suggested above. 

\medskip

\rem{2.2} In addition to the possible changes suggested in Remark 2.1, one can also incorporate dependence into our model. Recall that we assume the times of death $\tau_x$ and $\tau_y$ are independent in this paper; however, as a rule, the lives of people in a household are dependent (Frees et al., 1986).  Another sort of dependence that would be natural to include is time dependence.  Indeed, one's force of mortality generally increases with age (which would affect $\la_x$, $\la_y$, and $H$), and one's rate of income might be larger before retirement than afterwards.  We do not believe that introducing these dependences would materially change the results of this paper, although the mathematics might not be as tractable.

\medskip

\rem{2.3} After the first death, the remaining family member, either $(x)$ or $(y)$, faces a Merton problem of maximizing utility of lifetime consumption.  Specifically, the corresponding value function for the survivor, as a function of the hazard rate $\la$ and rate of income $I$ of the survivor, is given by
$$
V(w; \la, I) = \sup_{(\c, \Pi)} \E^w \left[ \int_0^{\tau_2} e^{-r t} \, u(c_t) \, dt \right] = \sup_{(\c, \Pi)} \E^w \left[ \int_0^\infty e^{-(r + \la) t} \, u(c_t) \, dt \right] ,
$$
in which wealth follows
$$
dW_t = (r W_t + (\mu - r) \pi_t + I  - c_t) \, dt + \sigma \, \pi_t \, dB_t, \quad W_0 = w.
$$
We know from Merton (1969, Section 9) that $V$ is given by
$$
V(w; \la, I) = - \, {1 \over \al r} \, \exp \left\{ - \al r \left( w + {I \over r} + {\la + m \over \al \, r^2} \right) \right\},
\eqno(2.5)
$$
in  which $m = {1 \over 2} \, \left( {\mu - r \over \sigma} \right)^2$.  Note that wealth in equation (2.5) is augmented by the present value of a perpetuity that pays at the rate of $I$, in addition to a term that accounts for the force of mortality and the financial market.  The optimal rate of consumption for the household after the first death is
$$
c^*_V(w; \la, I) = (u')^{-1} (V_w(w)) = r w + I + {\la + m \over \al \, r},
\eqno(2.6)
$$
which is the investment income if one were to invest all wealth in the riskless asset ($rw$), plus the survivor's earned income ($I$), plus an additional amount based on the survivor's force of mortality ($\la$), the riskiness of the market ($m$), and the risk aversion ($\al$).  The optimal amount to invest in the risky asset after the first death is
$$
\pi^*_V(w; \la, I) = - \, {\mu - r \over \sigma^2} \, {V_w(w) \over V_{ww}(w)} = {\mu - r \over \al \, r \, \sigma^2},
\eqno(2.7)
$$
a constant, which is the usual result when considering exponential utility.  Not only is $\pi^*_V$ a constant, it is independent of the force of mortality.

\rem{2.4} It follows from the Dynamic Programming Principle (see, for example, Jeanblanc et al.\ (2004, Proposition 2.4)) that we can rewrite $U$ in (2.3) as follows:
$$
\eqalign{
&U(w, D) = \sup_{(\c, \Pi, \D)} \E^{w, D} \left[ {\bf 1}_{\{ \tau_x < \tau_y \}} \left( \int_0^{\tau_x} e^{- r t} \, u(c_t) \, dt + e^{-r \tau_x} V \left( W_{\tau_x-} + D_{\tau_x-}; \ly, I_y \right) \right) \right. \cr
& \qquad \qquad \qquad \qquad \qquad \quad \left. + {\bf 1}_{\{ \tau_x \ge \tau_y \}} \left( \int_0^{\tau_y} e^{- r t} \, u(c_t) \, dt + e^{- r \tau_y} V \left( W_{\tau_y-} + D_{\tau_y-}; \lx, I_x \right) \right) \right], \cr
& \quad = \sup_{(\c, \Pi, \D)} \E^{w, D} \left[   \int_0^\infty e^{- \tilde \delta t} \left[ u(c_t) + \lx V(W_{t-} + D_{t-}; \ly, I_y) + \ly V(W_{t-} + D_{t-}; \lx, I_x) \right] dt  \right] ,
}
\eqno(2.8)
$$
in which $\tilde \delta \equiv r + \lx + \ly$, $u$ is given in (2.4), and $V$ is given in (2.5).  The first expectation includes the randomness in the times of death, while the second does not because the random times of death are accounted for within the integral.

\subsect{2.3. Verification lemma}

In this section, we provide a verification lemma that states that a smooth solution to a variational inequality associated with the maximization problem in (2.3) is the value function $U$.  Therefore, we can reduce our problem to one of solving a variational inequality.  We state the verification lemma without proof because its proof is similar to others in the literature; see, for example, Wang and Young (2012a, 2012b).

First, for $c, \pi \in \R$, define a differential operator $\L^{c, \pi}$ by its action on a test function $f$ on $\R \times \R^+$; the definition of the operator is derived from the wealth dynamics before $\tau_1$ in (2.2) and the alternative expression for $U$ in (2.8).
$$
\eqalign{
\L^{c, \pi} \, f(w, D) &= (rw + (\mu - r) \pi + I_x + I_y - c) f_w(w, D) + {1 \over 2} \sigma^2 \pi^2 f_{ww}(w, D) - {1 \over \al} e^{-\al c} \cr
& \quad - (r + \lx + \ly) f(w, D) + \lx \, V(w + D; \ly, I_y) + \ly \, V(w + D; \lx, I_x).
}
\eqno(2.9)
$$


\lem{2.1}{Let $\up = \up(w, D)$ be a function that is non-decreasing, concave, and twice-differentiable with respect to $w$ and non-decreasing and differentiable with respect to $D$ on $\Do$.  Suppose that $\up$ satisfies the following variational inequality:
$$
\max( \max_{c, \pi} \L^{c, \pi} \, \up(w, D), \, \up_D(w, D) - H \, \up_w(w, D)) = 0.
\eqno(2.10)
$$
\hfill \break
\noi Then, on $\Do$,
$$
U = \up.
$$
To specify the optimal strategies, first partition the region $\Do$ into ${\cal R}_1 := \{ (w, D) \in \Do: U_D(w, D) - H \, U_w(w, D) < 0 \}$ and ${\cal R}_2 := \{ (w, D) \in \Do:  U_D(w, D) - H \, U_w(w, D) = 0 \}$. Then, the optimal life insurance purchasing, consumption, and investment strategies are as follows:
\item{$(a)$} When $(W_{t-}, D_{t-}) \in {\cal R}_2$, purchase additional life insurance $\Delta D$ so that $(W_{t-} - H \Delta D, D_{t-} + \Delta D) \in \partial({\cal R}_1)$.
\item{$(b)$} When $(W_t, D_t) \in cl({\cal R}_1)$, purchase additional life insurance instantaneously to keep $(W_t, D_t) \in cl({\cal R}_1)$.
\item{$(c)$} Consume continuously at the following rate when $(W_t, D_t) = (w, D) \in cl({\cal R}_1)$:
$$
c^*_U(w, D) = (u')^{-1} (U_w(w, D)).
\eqno(2.11)
$$
\item{$(d)$} Invest the following amount of wealth in the risky asset when  $(W_t, D_t) = (w, D) \in cl({\cal R}_1)$:
$$
\pi^*_U(w, D) = - \, {\mu - r \over \sigma^2} \, {U_w(w, D) \over U_{ww}(w, D)}.
\eqno(2.12)
$$}

The region ${\cal R}_1$ is called the {\it continuation} region because when the wealth and life insurance benefit lie in the interior of the region, the household does not purchase more life insurance; it continues with its current benefit.  Indeed, $U_D < H \, U_w$ means that the marginal benefit of buying more life insurance ($U_D$) is less than the marginal cost of doing so ($H \, U_w$).  On the closure of that region, written $cl({\cal R}_1)$, the Hamilton-Jacobi-Bellman (HJB) equation holds:
$$
\max_{c, \pi} \L^{c, \pi} \, U(w, D) = 0.
\eqno(2.13)
$$

At the boundary of the continuation region, the household exercises singular control to remain in $cl({\cal R}_1)$. If the wealth and insurance benefit lie in ${\cal R}_2$, then the household uses impulse control to move to the boundary of ${\cal R}_1$, written $\partial({\cal R}_1)$. Because the household moves instantaneously from ${\cal R}_2$ to $\partial({\cal R}_1)$, we do not need to specify consumption and investment strategies on ${\cal R}_2$. Also, because of this instantaneous movement, if $(w, D) \in {\cal R}_2$, then
$$
U(w, D) = U(w - H \, \Delta D, D + \Delta D),
$$
in which $\Delta D$ is such that $(w - H \, \Delta D, D + \Delta D) \in \partial({\cal R}_1)$, so that the right-hand side is given by the solution to the HJB equation.

\sect{3. Insurance Purchased with a Single Premium}

Throughout this section, we assume that the household purchases its life insurance via single premiums $H = (1 + \theta) {\lx + \ly \over \lx + \ly + r}$ per dollar of insurance benefit, for some $\theta \ge 0$. In the next section, we consider the case for which the household pays its life insurance premium continuously until $\tau_1$.

\subsect{3.1. Maximized utility}

On the closure of the continuation region, $U = U(w, D)$ solves the HJB equation (2.13):
$$
\eqalign{
(r + \lx + \ly) U &= (rw + I_x + I_y) U_w + \max_c \left[ - \, {1 \over \al} \, e^{- \al c} - c U_w \right] + \max_\pi \left[ (\mu - r) \pi U_w + {1 \over 2} \sigma^2 \pi^2 U_{ww} \right] \cr
& \quad - {1 \over \al r} \, e^{- \al r (w + D) - m/r} \left[ \lx \, e^{- \al I_y - \ly/r} + \ly \, e^{- \al I_x - \lx/r} \right].
}
\eqno(3.1)
$$
Suppose that $U$ is of the form $U(w, D) = -  {k(D) \over \al r} \, e^{- \al r w}$; then, one can show $k(D)$ is given implicitly as the unique positive solution of the following equation:
$$
k \left[ r \ln k + \al r(I_x + I_y) + \lx + \ly + m \right] = e^{- \al r D - m/r} \left[ \lx \, e^{- \al I_y - \ly/r} + \ly \, e^{- \al I_x - \lx/r} \right].
\eqno(3.2)
$$
It will prove useful to know properties of $k(D)$.

\lem{3.1} {$k(D)$, defined as the unique positive solution of $(3.2)$, is decreasing with respect to $D$, while $k(D) e^{\al r D}$ is increasing and convex with respect to $D$.}

\pf  To show that $k'(D)$ decreases with respect to $D$, differentiate equation (3.2) to get
$$
\eqalign{
k'(D) &\left[ r \ln k(D) + \al r(I_x + I_y) + \lx + \ly + m + r \right]  \cr
&= - \al r \, e^{- \al r D - m/r} \left[ \lx \, e^{- \al I_y - \ly/r} + \ly \, e^{- \al I_x - \lx/r} \right].}
\eqno(3.3)
$$
It follows from equation (3.2) that the expression in the square brackets on the left-hand side of (3.3) is positive; thus, $k'(D) < 0$.

Next, from (3.2) and (3.3), observe that
$$
{d \over dD} \left(k(D) e^{\al r D} \right) \left[ r \ln k(D) + \al r(I_x + I_y) + \lx + \ly + m + r \right] = \al r^2 k(D) e^{\al r D},
$$
from which we deduce that the derivative of $k(D) e^{\al rD}$ with respect to $D$ is positive. Differentiate this expression with respect to $D$ to obtain
$$
\eqalign{
{d^2 \over dD^2} \left(k(D) e^{\al r D} \right) &\left[ r \ln k(D) + \al r(I_x + I_y) + \lx + \ly + m + r \right] + {d \over dD} \left(k(D) e^{\al r D} \right) {r k'(D) \over k(D)} \cr
&= \al r^2 {d \over dD} \left(k(D) e^{\al r D} \right),}
$$
which implies that the second derivative of $k(D) e^{\al r D}$ with respect to $D$ is positive.  \qed

Thus, the function $(w, D) \to - {k(D) \over \al r} e^{- \al r w}$ is increasing, concave with respect to $w$ and increasing with respect to $D$.  According to Lemma 2.1, it is a good candidate for our maximized utility because it also solves the HJB equation (3.1).

In the interior of the continuation region, we have $U_D(w, D) - H \, U_w(w, D) < 0$, or equivalently,
$$
k(D) < \exp \left\{ {H \over 1 - H} - \al (I_x + I_y) - {\lx + \ly + m \over r} \right\}.
\eqno(3.4)
$$
Note that the region defined by (3.4) is independent of wealth $w$, so in the case for exponential utility, the resulting control problem is {\it not} one of stochastic control.  Either the household has enough life insurance $D$ so that inequality (3.4) holds weakly, or the household buys enough insurance so that (3.4) holds with equality.

In mathematical terms, if $k(D)$ is greater than the expression on the right-hand side of inequality (3.4), then we are in the interior of the ``buy'' region ${\cal R}_2$, and the household will buy enough insurance (increase $D$) so that we have equality in (3.4). Recall that $k'(D) < 0$, so $k(D)$ decreases as the household buys more insurance.

We next find an expression for the boundary between the continuation and the buy regions, ${\cal R}_1$ and ${\cal R}_2$, respectively.  In the case for exponential utility, this boundary is a line of the form $\{ (w, D^*): w \in \R \}$ for a fixed value of $D^* \ge 0$. Despite the fact that the target death benefit, $D^*$, is independent of wealth, it still depends on the parameters of the model in interesting ways.

For $D = D^*$, we have equality in (3.4), which gives us an explicit expression for $k(D^*)$. We, then, substitute this expression into (3.2) and solve for the optimal amount of life insurance $D^*$:
$$
D^* = \max \left[ {1 \over \al r} \left[ \ln \left( \lx \exp \left(\al I_x + {\lx \over r} \right) + \ly \exp \left(\al I_y + {\ly \over r} \right) \right) - \ln \left( {rH \over 1- H} \right) - {H \over 1 - H} \right], \, 0 \, \right].
\eqno(3.5)
$$
Note that we do not allow the household to buy a negative amount of life insurance; thus, we force $D^* \ge 0$ in the expression in (3.5).  In the next proposition, we give the maximized utility in the case for which $D^* = 0$ in (3.5). In this case, the continuation region ${\cal R}_1$ equals the entire domain $\R \times \R^+$.

\prop{3.2} {If $D^* = 0$ in $(3.5)$, then the maximized utility $U$ defined by $(2.3)$ on $\R \times \R^+$ is given by:
$$
U(w, D) = - {k(D) \over \al r} \, e^{- \al r w},
\eqno(3.6)
$$
in which $k(D) > 0$ uniquely solves $(3.2)$. The associated optimal life insurance purchasing, consumption, and investment strategies are as follows:
\item{$(a)$} Do not purchase additional life insurance.
\item{$(b)$} Consume continuously at the following rate when $(W_t, D_t) = (w, D) \in \R \times \R^+:$
$$
c^*_U(w, D) = r w - {1 \over \al} \, \ln k(D).
\eqno(3.7)
$$
\item{$(c)$} Invest the following amount of wealth in the risky asset when $(W_t, D_t) = (w, D) \in \R \times \R^+:$
$$
\pi^*_U(w, D) =  {\mu - r \over \al \, r \, \sigma^2}.
\eqno(3.8)
$$}

\pf We use Lemma 2.1 to prove this proposition. Because $U$ in (3.6) solves the HJB equation (3.1), to verify that equation (2.10) holds, it is enough to show that $U_D(w, D) - H \, U_w(w, D) \le 0$ on $\R \times \R^+$, which is equivalent to showing that (3.4) holding weakly.  The left-hand side of (3.2) increases with respect to $k$; thus, to show that (3.4) holds, it is enough to show that the equality in (3.2) is replaced by $\ge$ when we replace $k$ with the right-hand side of (3.4).  Specifically, it is enough to show that
$$
{r H \over 1 - H} \, \exp \left[ {H \over 1 - H} - \al (I_x + I_y) - {\lx + \ly + m \over r} \right] \ge e^{- \al r D - m/r} \left[ \lx \, e^{- \al I_y - \ly/r} + \ly \, e^{- \al I_x - \lx/r} \right] ,
$$
for all $D \in \R^+$, or equivalently,
$$
{r H \over 1 - H} \, \exp \left( {H \over 1 - H} \right) \ge \lx \, e^{ \al I_x + \lx/r} + \ly \, e^{ \al I_y + \ly/r},
$$
which holds  if and only if $D^* = 0$ in (3.5). Thus, we have verified that $U$ in equation (3.6) solves the variational inequality (2.10).  Because the entire space is the continuation region, it is optimal {\it not} to purchase more life insurance. Finally, the optimal consumption and investment strategies are given in feedback form by the first-order necessary conditions of the HJB equation, as in equations (2.11) and (2.12), respectively.  \qed

Next, we consider the case for which $D^* > 0$ in (3.5), and we state the main theorem of this section. Then, in Section 3.2, we examine how $D^*$ varies with the parameters of the model.

\thm{3.3} {If $D^* > 0$ in $(3.5)$, then the maximized utility $U$ defined by $(2.3)$ on $\R \times \R^+ = {\cal R}_1 \cup {\cal R}_2$, with ${\cal R}_1 = \{ (w, D): w \in \R, \, D > D^* \}$ and ${\cal R}_2 = \{ (w, D): w \in \R, \, 0 \le D \le D^* \}$, is given by:
\item{$(i)$} For $(w, D) \in cl({\cal R}_1) = \{ (w, D): w \in \R, \, D \ge D^* \}$,
$$
U(w, D) = - {k(D) \over \al r} \, e^{- \al r w}, 
\eqno(3.9)
$$
in which $k(D) > 0$ uniquely solves $(3.2)$.
\item{$(ii)$} For $(w, D) \in {\cal R}_2$,
$$
U(w, D) = U(w - H (D^* - D), D^*).
\eqno(3.10)
$$
\indent The associated optimal life insurance purchasing, consumption, and investment strategies are as follows:
\item{$(a)$} Purchase additional life insurance of $\Delta D = D^* - D$ when $D_{t-} = D < D^*$.
\item{$(b)$} Do not purchase additional life insurance when $D_{t-} = D \ge D^*$.
\item{$(c)$} Consume continuously at the rate given by $(3.7)$ when $(W_t, D_t) = (w, D) \in cl({\cal R}_1)$.
\item{$(d)$} Invest the amount of wealth in the risky asset given by $(3.8)$ when $(W_t, D_t) = (w, D) \in cl({\cal R}_1)$.}

\pf As in the proof of Proposition 3.2, we use Lemma 2.1 to prove this theorem.  By construction, $U$ in (3.9) satisfies the HJB equation on $cl({\cal R}_1)$.  We next show that $U$ in (3.10) satisfies $\L^{c, \pi} \, U \le 0$ on ${\cal R}_2$ for all $c, \pi \in \R$.  To this end, observe that for $(w, D) \in {\cal R}_2$, $U_w(w, D) = U_w(w - H \, \Delta D, D^*)$ and similarly for the second derivative with respect to $w$.  Thus, for $(w, D) \in {\cal R}_2$, we have from the definition of $\L^{c, \pi}$ in (2.9),
$$
\eqalign{
& \L^{c, \pi} \, U(w, D)  = (rw + (\mu - r) \pi + I_x + I_y - c) U_w(w, D) + {1 \over 2} \sigma^2 \pi^2 U_{ww}(w, D) + u(c) \cr
& \qquad \qquad \qquad \quad - (r + \lx + \ly) U(w, D) + \lx \, V(w + D; \ly, I_y) + \ly \, V(w + D; \lx, I_x) \cr
&= (r(w - H \, \Delta D) + (\mu - r) \pi + I_x + I_y - c) U_w(w, D) + {1 \over 2} \sigma^2 \pi^2 U_{ww}(w, D) + u(c) \cr
& \quad - (r + \lx + \ly) U(w, D) + \lx \, V(w + D^*; \ly, I_y) + \ly \, V(w + D^*; \lx, I_x) \cr
& \quad + r H \, \Delta D \, U_w(w, D) + \lx (V(w + D; \ly, I_y) - V(w + D^*; \ly, I_y)) \cr
& \quad + \ly ( V(w + D; \lx, I_x) - V(w + D^*; \lx, I_x)) \cr
&= \L^{c, \pi} U(w - H \, \Delta D, D^*) + r H \, \Delta D \, U_w(w, D) \cr
& \quad + \lx (V(w + D; \ly, I_y) - V(w + D^*; \ly, I_y)) + \ly ( V(w + D; \lx, I_x) - V(w + D^*; \lx, I_x)) \cr
& \le r H \, \Delta D \, U_w(w - H \, \Delta D, D^*) + \lx (V(w + D; \ly, I_y) - V(w + D^*; \ly, I_y)) \cr
& \quad + \ly ( V(w + D; \lx, I_x) - V(w + D^*; \lx, I_x)) \cr
& = r H \, \Delta D \, \exp \left\{ {H \over 1 - H} - \al (I_x + I_y) - {\lx + \ly + m \over r} \right\} \, e^{- \al r (w - H \, \Delta D)} \cr
& \quad - {1 \over \al r} \, \exp \left( -\al r w - {m \over r} \right) \, \left(e^{- \al r D} - e^{- \al r D^*} \right) \left[ \lx \exp \left( - \al I_y - {\ly \over r} \right) + \ly \exp \left( -\al I_x - {\lx \over r} \right) \right],}
$$
in which the inequality follows from the fact that $\L^{c, \pi} \, U \le 0$ on $cl({\cal R}_1)$.  Also, in the last expression, we substitute for both $U$ and $V$ and use the fact that $k(D^*)$ for $D^* > 0$ is given by the right-hand side of (3.4).  After simplifying, we learn that the last expression is non-positive if and only if
$$
r H \, \Delta D \, e^{ \al r H \Delta D + {H \over 1- H}} \le {1 \over \al r} \, \left(e^{- \al r D} - e^{- \al r D^*} \right) \left[ \lx \exp \left( \al I_x + {\lx \over r} \right) + \ly \exp \left( \al I_y + {\ly \over r} \right) \right].
$$
From the definition of $D^* > 0$ in (3.5), substitute for $\left[ \lx \exp \left( \al I_x + {\lx \over r} \right) + \ly \exp \left( \al I_y + {\ly \over r} \right) \right]$ in the above to get the following inequality:
$$
r H \, \Delta D \, e^{\al r H \Delta D + {H \over 1- H}} \le {1 \over \al r} \, \left(e^{- \al r D} - e^{- \al r D^*} \right) e^{\al r D^*} \, {rH \over 1 - H} \, e^{{H \over 1 - H}},
$$
or equivalently,
$$
e^{- \al r \Delta D} + \al r (1 - H) \Delta D \, e^{- \al r (1 - H) \Delta D} \le 1,
$$
in which we rely on the assumption that $H < 1$. This last inequality holds if $e^{-a} + ab e^{-ab} \le 1$ for any $a > 0$ and $b \in (0, 1)$, which, via elementary analysis, one can show is true. Thus, we have demonstrated that $\max_{c, \pi} \L^{c, \pi} \, U \le 0$ on $\R \times \R^+$ with equality on $cl({\cal R}_1)$.

Next, we show that $U_D - H \, U_w \le 0$ on $\R \times \R^+$. By construction, this inequality holds with equality on ${\cal R}_2$; see equation (3.10).  Now, $U_D - H \, U_w < 0$ if and only if inequality (3.4) holds.  By definition of $D^*$, $k(D^*)$ is given by the right-hand side of (3.4). Recall that $k(D)$ strictly decreases with respect to $D$; thus, inequality (3.4) holds exactly when $D > D^*$, or equivalently on ${\cal R}_1$.   Thus, we have shown that $U_D - H \, U_w \le 0$ on $\R \times \R^+$ with strict inequality on ${\cal R}_1$ and equality on ${\cal R}_2$.

We have verified that $U$ given in equations (3.9) and (3.10) satisfies the variational inequality (2.10) in Lemma 2.1.  Also, $U$ is non-decreasing, concave, and twice-differentiable with respect to $w$ and non-decreasing and differentiable with respect to $D$.  Therefore, $U$ as stated is the maximized utility, and the optimal strategies follow from Lemma 2.1.  \qed

\rem{3.1} Because the optimal death benefit $D^*$ is independent of wealth, one can alternatively determine $D^*$ as follows:  First, suppose the household begins with wealth $w$ and no life insurance.  If the household then buys life insurance of amount $D$, the maximum utility equals $U(w - HD, D)$ in which $U$ is given by equation (3.6) or  (3.9).  Next, find $D = D^* \ge 0$ to maximize $U(w - HD, D)$, or equivalently, find $D^* \ge 0$ to minimize $k(D) e^{\al r H D}$.  Because $k(D) e^{\al r H D}$ is convex with respect to $D$, there is a unique value of $D^* \ge 0$ that minimizes it, and one can show that that value is the one given in (3.5).

\medskip

We have the following corollary to Proposition 3.2 and Theorem 3.3, in which we compare the optimal consumption and investment immediately before and after the first death.

\cor{3.4} {At the moment of the first death, the optimal rate of consumption changes by
$$
\Delta c(D) = r D + I_a + {\la_a + m \over \al r} + {1 \over \al} \ln k(D),
\eqno(3.11)
$$
in which $I_a$ and $\la_a$ are the income rate and force of mortality of the survivor, respectively.  When $D = D^* > 0$, then the expression in $(3.11)$ becomes
$$
\Delta c(D^*) = {1 \over \al} \left[ \ln \left( {\la_d \over \lx + \ly} + {\la_a \over \lx + \ly} \exp \left( \al(I_a - I_d) + {\la_a - \la_d \over r} \right) \right) + \ln \left( {r - \theta (\lx + \ly) \over r(1 + \theta)} \right) \right] ,
\eqno(3.12)
$$
in which the subscripts $a$ and $d$ refer to the one who is alive or dead, respectively.  The optimal amount invested in the risky asset does not change upon the first death.}

\pf  From Proposition 3.2 and Theorem 3.3, we know that the optimal rate of consumption immediately before the first death equals $r w - {1 \over \al} \ln k(D)$.  From equation (2.6), the optimal rate of consumption immediately after the first death equals $r(w + D) + I + (\la + m)/(\al r)$, in which $I$ and $\la$ are the income rate and force of mortality of the survivor, respectively.  The change in the rate of consumption, therefore, equals the expression in (3.11).

When $D^* > 0$, the right-hand side of inequality (3.4) equals $k(D^*)$ and (3.5) gives us $D^*$.  After substituting these expressions into (3.11), we obtain (3.12). From equations (2.7) and (3.8), we see that the optimal amount invested in the risky asset is constant, before and after the first death, independent of wealth and the death benefit.  \qed

Numerical work indicates that the change in consumption can be positive or negative. We deduce the following properties of the change in consumption from the expression in (3.11):

\smallskip

\item{(i)}  If the survivor has greater income and greater force of mortality than the household member who died, then the change in consumption is greater than if the latter had survived. This result makes sense because the surviving household member has more income from which to consume and has a shorter expected time until dying so will also consume more for that reason.

\item{(ii)} From (3.2) and (3.3), one can show that $\al r D + \ln k(D)$ increases with $D$; thus, $\Delta c(D)$ increases with $D$.  So, for a household with a larger death benefit, that household's change in consumption will be greater than for a household with a smaller death benefit. This result, too, makes sense because--all else equal--one expects that a household receiving a larger death benefit will consume more.

\item{(iii)} As a special case of observation (ii), if a household does not buy any life insurance, then the change in the rate of consumption at the first death is $\Delta c(0)$, which is {\it less} than $\Delta c(D)$ for any $D > 0$.

\subsect{3.2. Properties of the optimal death benefit}

In this section, we determine how $D^*$ varies with the parameters of the model.  Most results are as one might expect; however, one of them is rather surprising. Indeed, in Corollary 3.7, we find that if the household is close to risk neutral but still risk averse ($\al > 0$ but small), then buying no life insurance is optimal, even when it is priced fairly ($\theta = 0$).  We also learn in Proposition 3.10 that the optimal amount of insurance is bounded above by the maximal loss, as measured by the present value of the perpetuity that pays at the rate of $\max(I_x, I_y)$, an intuitively pleasing result.

First, note that although the optimal consumption and investment strategies depend on the drift and volatility of the risky asset, the optimal death benefit $D^*$ does not.

\prop{3.5} {The optimal amount of life insurance $D^*$ decreases with the premium loading $\theta$ to the extent that for $\theta$ large enough, buying no life insurance is optimal.}

\pf From the expression for $D^*$ in (3.5), define $f$ by
$$
f(\theta) =  - \ln \left( {rH(\theta) \over 1- H(\theta)} \right) - {H(\theta) \over 1 - H(\theta)},
$$
in which $H = H(\theta)$ is given in (2.1). It is enough to show that $f$ decreases with $\theta$.
$$
f'(\theta) = - {H'(\theta) \over H(\theta) (1 - H(\theta))^2} < 0,
$$
because $H'(\theta) > 0$.  From (3.5), we see that as $H$ approaches 1, $D^*$ goes to 0. \qed

It is no surprise that the optimal amount death benefit decreases as the loading on the single premium increases.  Proposition 3.5 helps to confirm the reasonableness of our model's qualitative results.  Similarly, we expect that the optimal death benefit increases as the household becomes more risk averse, and this is indeed the case, which we prove in the next proposition.

\prop{3.6} {The optimal amount of life insurance $D^*$ increases with the risk aversion of the household, as measured by $\al$.}

\pf From (3.5), define $g$ by
$$
g(\al) =  {1 \over \al} \left[ \ln \left( \lx \exp \left(\al I_x + {\lx \over r} \right) + \ly \exp \left(\al I_y + {\ly \over r} \right) \right) - \ln \left( {rH \over 1- H} \right) - {H \over 1 - H} \right];
\eqno(3.13)
$$
then,
$$
\eqalign{
g'(\al) &=  - {1 \over \al^2} \left[ \ln \left( \lx \exp \left(\al I_x + {\lx \over r} \right) + \ly \exp \left(\al I_y + {\ly \over r} \right) \right) - \ln \left( {rH \over 1- H} \right) - {H \over 1 - H} \right] \cr
& \quad + {1 \over \al} \, {I_x \lx \exp \left(\al I_x + {\lx \over r} \right) + I_y \ly \exp \left(\al I_y + {\ly \over r} \right) \over  \lx \exp \left(\al I_x + {\lx \over r} \right) + \ly \exp \left(\al I_y + {\ly \over r} \right)},}
$$
which is positive if and only if
$$
\eqalign{
& \ln \left( \lx  \exp \left(\al I_x + {\lx \over r} \right) + \ly  \exp \left(\al I_y + {\ly \over r} \right) \right) - \ln \left( {rH \over 1 - H} \right) - {H \over 1 - H} \cr
& \quad < \al \, {I_x \lx \exp \left(\al I_x + {\lx \over r} \right) + I_y \ly \exp \left(\al I_y + {\ly \over r} \right) \over  \lx \exp \left(\al I_x + {\lx \over r} \right) + \ly \exp \left(\al I_y + {\ly \over r} \right)}.}
\eqno(3.14)
$$
When $\al = 0$, one can show that the left-hand side of (3.14) is negative. Indeed, when $\theta = 0$ and $\al = 0$, it reduces to $\ln \left( {\lx \over \lx + \ly}  e^{- \ly/r} + {\ly \over \lx + \ly} e^{- \lx/r} \right) < 0$. Because the left-hand side of (3.14) decreases with respect to $\theta$ when $H < 1$, it is negative for all $\theta \ge 0$ such that $H < 1$.

Because inequality (3.14) holds when $\al = 0$, it is enough to show that the derivative of the left-hand side with respect to $\al$ is less than or equal to the derivative of the right-hand side. This ordering of the derivatives is equivalent to
$$
{d \over d \al} {I_x \lx \exp \left(\al I_x + {\lx \over r} \right) + I_y \ly \exp \left(\al I_y + {\ly \over r} \right) \over  \lx \exp \left(\al I_x + {\lx \over r} \right) + \ly \exp \left(\al I_y + {\ly \over r} \right)} \ge 0,
$$
which is straightforward to demonstrate.  \qed

From the proof of Proposition 3.6, we have the following corollary.

\cor{3.7} {For $\theta \ge 0$, there exists $\al(\theta) > 0$ such that purchasing no insurance is optimal if and only if $\al \le \al(\theta)$.}

In other words, even for actuarially fair insurance (that is, $\theta = 0$), if the household is not very risk averse, then it will not buy life insurance. This result contrasts dramatically with the one in, say, casualty insurance in which a risk averse agent will buy {\it full} coverage for a loss if insurance is priced fairly; see, for example, Exercise 1.22 in Bowers et al.\ (1997).

In the next proposition, we explore when the optimal amount of life insurance is concave with respect to $\al$.  That is, as $\al$ increases, the optimal death benefit increases (as we know from Proposition 3.6) but at an increasingly slower rate.

\prop{3.8} {If $(I_x, \lx) \ge (I_y, \ly)$ or if $(I_x, \lx) \le (I_y, \ly)$, then the optimal amount of life insurance $D^*$ is concave with respect to the risk aversion of the household, as measured by $\al$, when $D^* > 0$.}

\pf It is enough to show that $g$ in (3.13) has a negative second derivative with respect to $\al$.
$$
\eqalign{
g''(\al) &=  {2 \over \al^3} \left[ \ln \left( \lx \exp \left(\al I_x + {\lx \over r} \right) + \ly \exp \left(\al I_y + {\ly \over r} \right) \right) - \ln \left( {rH \over 1- H} \right) - {H \over 1 - H} \right] \cr
& \quad - {2 \over \al^2} \, {I_x \lx \exp \left(\al I_x + {\lx \over r} \right) + I_y \ly \exp \left(\al I_y + {\ly \over r} \right) \over  \lx \exp \left(\al I_x + {\lx \over r} \right) + \ly \exp \left(\al I_y + {\ly \over r} \right)} \cr
& \quad + {1 \over \al} \, {(I_x - I_y)^2 \lx \ly \exp \left( \al (I_x + I_y) + {\lx + \ly \over r} \right) \over \left( \lx \exp \left(\al I_x + {\lx \over r} \right) + \ly \exp \left(\al I_y + {\ly \over r} \right) \right)^2},
}
$$
which is negative if and only if
$$
\eqalign{
& \ln \left( \lx \exp \left(\al I_x + {\lx \over r} \right) + \ly \exp \left(\al I_y + {\ly \over r} \right) \right) - \ln \left( {rH \over 1- H} \right) - {H \over 1 - H} \cr
& < \al \, {I_x \lx \exp \left(\al I_x + {\lx \over r} \right) + I_y \ly \exp \left(\al I_y + {\ly \over r} \right) \over  \lx \exp \left(\al I_x + {\lx \over r} \right) + \ly \exp \left(\al I_y + {\ly \over r} \right)} - {\al^2 \over 2} \, {(I_x - I_y)^2 \lx \ly \exp \left( \al (I_x + I_y) + {\lx + \ly \over r} \right) \over \left( \lx \exp \left(\al I_x + {\lx \over r} \right) + \ly \exp \left(\al I_y + {\ly \over r} \right) \right)^2}.}
\eqno(3.15)
$$
As we observed in the proof of Proposition 3.6, the left-hand side is negative when $\al = 0$; thus, it is enough to show that the derivatives with respect to $\al$ of the two sides of (3.15) are ordered.  This ordering of the derivatives is equivalent to
$$
{d \over d \al} {(I_x - I_y)^2 \lx \ly \exp \left( \al (I_x + I_y) + {\lx + \ly \over r} \right) \over \left( \lx \exp \left(\al I_x + {\lx \over r} \right) + \ly \exp \left(\al I_y + {\ly \over r} \right) \right)^2} \le 0,
$$
which reduces to
$$
(I_x - I_y) \left(  \lx \exp \left(\al I_x + {\lx \over r} \right) - \ly \exp \left(\al I_y + {\ly \over r} \right)  \right) \ge 0.
$$
This last inequality holds if either $(I_x, \lx) \ge (I_y, \ly)$ or if $(I_x, \lx) \le (I_y, \ly)$, and we are done. \qed

Even though ordering of the derivatives in (3.15) is a stronger condition than one needs to prove that $g$ is concave with respect to $\al$, numerical experiments indicate that $g$ might not be concave when the conditions of Proposition 3.8 do not hold.

Next, we observe that $D^*$ is increasing and convex with respect to the incomes of the household; we state this result without proof because it is clear from the expression for $D^*$ in (3.5).

\prop{3.9} {When $D^* > 0$, the optimal amount of life insurance is increasing and convex with respect to $I_x$ and $I_y$.}

The sign of the derivative of $D^*$ with respect to either force of mortality is ambiguous.  Indeed, from the expression for $D^*$ in (3.5), we deduce that increasing $\lx$ or $\ly$ shortens the expected horizon, which tends to increase $D^*$ via the logarithmic term; on the other hand, increasing either force of mortality increases the cost of life insurance, which tends to decrease $D^*$ via the $H$-terms.

We end this section with an upper bound on the life insurance purchased by a household.

\prop{3.10} {$$
D^* \le {\max (I_x, I_y) \over r} = \max(I_x, I_y) \, \bar a_{\overline{\infty}|}.
$$ That is, $D^*$ is bounded above by the present value of a perpetuity that pays at the rate of $\max(I_x, I_y)$. Moreover, $\lim_{\al \to \infty} D^* = \max(I_x, I_y) \, \bar a_{\overline{\infty}|}$.}

\pf From Proposition 3.5, we know that $D^*$ for $\theta \ge 0$ is bounded above by the $D^*$ corresponding to $\theta = 0$. Thus, from (3.5), we have
$$
\eqalign{
D^* & \le \max \left[ {1 \over \al r} \left[ \ln \left( {\lx \over \lx + \ly} \exp \left(\al I_x + {\lx \over r} \right) + {\ly \over \lx + \ly} \exp \left(\al I_y + {\ly \over r} \right) \right) - {\lx + \ly \over r} \right], \, 0 \, \right] \cr
& \le  \max \left[ {1 \over \al r} \left[ \al \, \max(I_x, I_y) + \ln \left( {\lx \over \lx + \ly} e^{- \ly/r} + {\ly \over \lx + \ly} e^{- \lx/r}  \right) \right], \, 0 \, \right] \cr
& <  {1 \over \al r} \left( \al \, \max(I_x, I_y) \right)  =  {\max (I_x, I_y) \over r}.
}
$$
By applying L'H\^opital's rule to the expression for $D^*$ in (3.5), we obtain $\lim_{\al \to \infty} D^* = \max(I_x, I_y) \, \bar a_{\overline{\infty}|}$. \qed

It is interesting that the optimal amount of death benefit is bounded above by the maximal ``loss'' that the household might sustain, in which we view the loss as the present value of the income stream.  Thus, it is never optimal for the household to buy ``too much'' insurance. 

\sect{4. Insurance Purchased by a Continuously Paid Premium}

In this section, we modify the work of Sections 2 and 3 by assuming that the household buys its life insurance via a premium paid continuously until the first death at the rate of $h = (1 + \bt) (\lx + \ly)$ per dollar of insurance for some $\bt \ge 0$. 

\subsect{4.1. Maximized utility}

As in Section 3.1, the optimal strategy for purchasing life insurance is for the household to buy additional insurance of $\overline D^* - D$ if $D < \overline D^*$ for some optimal amount of insurance $\bD$; otherwise, the household buys no further insurance.  In this section, we demonstrate this result and provide an expression for $\bD$.

With continuously paid premium for life insurance, wealth follows the dynamics
$$
\left\{
\eqalign{
dW_t &= (r W_{t-} + (\mu - r) \pi_{t-} + I_x + I_y  - c_{t-} - h D_{t-}) \, dt + \sigma \, \pi_{t-} \, dB_t, \quad 0 \le t < \tau_1, \cr
W_{\tau_1} &= W_{\tau_1-} + D_{\tau_1-} \, , \cr
dW_t &= (r W_t + (\mu - r) \pi_t + I_x \, {\bf 1}_{\{\tau_1 = \tau_y \}}  + I_y \, {\bf 1}_{\{\tau_1 = \tau_x \}}  - c_t) \, dt + \sigma \, \pi_t \, dB_t, \quad \tau_1 < t < \tau_2.
}
\right.
\eqno(4.1)
$$
By comparison with the wealth process in (2.2), note that the only change is to replace the single-premium term $-H dD_t$ with the continuously-paid-premium term $-h D_{t-} dt$.

Denote the maximized utility function by $\overline U$, which is defined as in equation (2.3) with the wealth dynamics given in (4.1).  Its HJB equation when $D \ge \overline D^*$ is similar to the one for $U$ given in (3.1) with an additional term on the right-hand side equal to $- h D U_w$. Thus, on $\R \times \R^+$, $\overline U$ solves the variational inequality:
$$
\max( \max_{c, \pi} \L^{c, \pi} \, \bU(w, D) - h D \bU_w(w, D), \, \bU_D(w, D)) = 0.
\eqno(4.2)
$$
The analog of the verification lemma, Lemma 2.1, holds in this case.

If we hypothesize a solution of the form $\bU(w, D) = - {\bk(D) \over \al r} e^{-\al r w}$, then $\bk(D)$ is given implicitly as the unique positive solution of the following equation:
$$
\bk \left[ r \ln \bk - \al r h D + \al r(I_x + I_y) + \lx + \ly + m \right] = e^{- \al r D - m/r} \left[ \lx \, e^{- \al I_y - \ly/r} + \ly \, e^{- \al I_x - \lx/r} \right].
\eqno(4.3)
$$
Note equation (4.3)'s similarity to (3.2).  


On the continuation region, we have $\bU_D < 0$ because on that region it is optimal not to purchase life insurance; that is, by increasing $D$, we would decrease $\bU$.  Note that $\bU_D < 0$ is equivalent to $\bk'(D) > 0$, which one can show occurs if and only if $D > \bD$, in which $\bD$ is given by
$$
\bD = \max \left[ {1 \over \al (h + r)} \left[ \ln \left( \lx \exp \left(\al I_x + {\lx \over r} \right) + \ly \exp \left(\al I_y + {\ly \over r} \right) \right) - \ln h - {h \over r} \right], \, 0 \, \right],
\eqno(4.4)
$$
the analog of $D^*$ in (3.5). In the case for which $\theta = 0 = \overline \theta$, we have ${H \over 1 - H} = {h \over r}$, so in that case, the only difference between (3.5) and (4.4) is the divisor of $\al r$ versus $\al (h + r) = \al (\lx + \ly + r)$, respectively. Note that $1/r = \bar a_{\overline{\infty}|}$, while $1/(\lx + \ly + r) = \bar a_{xy}$.

As in Section 3.1, we have two cases to consider: (1) $\bD = 0$, and (2) $\bD > 0$.  The next two results are the analogs of Proposition 3.2 and Theorem 3.3, respectively.  Their proofs are similar to the ones in Section 3.1, so we omit them.

\prop{4.1} {If $\bD = 0$ in $(4.4)$, then the maximized utility of lifetime consumption $\bU$ on $\R \times \R^+$ is given by:
$$
\bU(w, D) = - {\bk(D) \over \al r} \, e^{- \al r w},
\eqno(4.5)
$$
in which $\bk(D) > 0$ uniquely solves $(4.3)$. The associated optimal life insurance purchasing, consumption, and investment strategies are as follows:
\item{$(a)$} Do not purchase additional life insurance.
\item{$(b)$} Consume continuously at the following rate when $(W_t, D_t) = (w, D) \in \R \times \R^+:$
$$
c^*_{\bU}(w, D) = r w - {1 \over \al} \, \ln \bk(D).
\eqno(4.6)
$$
\item{$(c)$} Invest the following amount of wealth in the risky asset when $(W_t, D_t) = (w, D) \in \R \times \R^+:$
$$
\pi^*_{\bU}(w, D) =  {\mu - r \over \al \, r \, \sigma^2}.
\eqno(4.7)
$$}

\thm{4.2} {If $\bD > 0$ in $(4.4)$, then the maximized utility of consumption $\bU$ on $\R \times \R^+ = \bDo_1 \cup \bDo_2$, with $\bDo_1 = \{ (w, D): w \in \R, \, D > \bD \}$ and $\bDo_2 = \{ (w, D): w \in \R, \, 0 \le D \le \bD \}$, is given by:
\item{$(i)$} For $(w, D) \in cl(\bDo_1) = \{ (w, D): w \in \R, \, D \ge \bD \}$,
$$
\bU(w, D) = - {\bk(D) \over \al r} \, e^{- \al r w}, 
\eqno(4.8)
$$
in which $\bk(D) > 0$ uniquely solves $(4.3)$.
\item{$(ii)$} For $(w, D) \in \bDo_2$,
$$
\bU(w, D) = \bU(w, \bD).
\eqno(4.9)
$$
\indent The associated optimal life insurance purchasing, consumption, and investment strategies are as follows:
\item{$(a)$} Purchase additional life insurance of $\Delta D = \bD - D$ when $D_{t-} = D < \bD$.
\item{$(b)$} Do not purchase additional life insurance when $D_{t-} = D \ge \bD$.
\item{$(c)$} Consume continuously at the rate given by $(4.6)$ when $(W_t, D_t) = (w, D) \in cl(\bDo_1)$.
\item{$(d)$} Invest the amount of wealth in the risky asset given by $(4.7)$ when $(W_t, D_t) = (w, D) \in cl(\bDo_1)$.}

\rem{4.1} As in the case of a single premium, there is an alternative way to find the optimal death benefit $\bD$; see Remark 3.1.  Because the optimal death benefit is independent of wealth, one can find $\bD$ as follows:  First, for a given death benefit $D$, the maximum utility equals $\bU(w, D)$ in which $\bU$ is given by equation (4.5) or  (4.8).  Then, find $D = \bD \ge 0$ to maximize $\bU(w, D)$, or equivalently, find $\bD \ge 0$ to minimize $\bk(D)$.  Because $\bk(D)$ is convex, there is a unique value of $\bD \ge 0$ that minimizes it, and one can show that that value is the one given in (4.4).

\medskip

Parallel to Corollary 3.4, we have the following corollary to Proposition 4.2 and Theorem 4.2, in which we compare the optimal consumption and investment immediately before and after the first death.  We also compare this change with the corresponding change when premium is paid in a lump sum for a given death benefit. 

\cor{4.3} {At the moment of the first death, the optimal rate of consumption changes by
$$
\Delta \bc(D) = r D + I_a + {\la_a + m \over \al r} + {1 \over \al} \ln \bk(D),
$$
in which $I_a$ and $\la_a$ are the income rate and force of mortality of the survivor, respectively.  As before, the optimal amount invested in the risky asset does not change upon the first death.  Furthermore, for a given death benefit $D \ge 0$, $\Delta \bc(D) \ge \Delta c(D)$ with equality only when $D = 0$.}

\pf We rewrite equations (3.2) and (4.3), respectively:
$$
r k(D) e^{\al(I_x + I_y) + {\lx + \ly + m \over r}} \ln \left( k(D) e^{\al(I_x + I_y) + {\lx + \ly + m \over r}} \right) = e^{-\al r D} \left( \lx e^{\al I_x + {\lx \over r}} + \ly e^{\al I_y + {\ly \over r}} \right),
$$
and
$$
r \bk(D) e^{\al(I_x + I_y) + {\lx + \ly + m \over r}} \left[ \ln \left( \bk(D) e^{\al(I_x + I_y) + {\lx + \ly + m \over r}} \right) - \al h D \right] = e^{-\al r D} \left( \lx e^{\al I_x + {\lx \over r}} + \ly e^{\al I_y + {\ly \over r}} \right).
$$
Thus, $\bk(D) \ge k(D)$ with equality only when $D = 0$, which completes our proof.  \qed

That the rate of change of consumption is greater in the case of continuously paid premium is not a surprise because part of the outflow of this household before the time of the first death is the premium, so that the first death not only triggers receipt of the death benefit $D$, it triggers the end of the premium payment.

\subsect{4.2. Properties of the optimal death benefit}

In this section, we determine how $\bD$ varies with the parameters of the model.  Most of the results of Section 3.2 hold in this case because the expression for $\bD$ is similar to one for $D^*$, including the surprising Corollary 3.7, in which we found that a risk-averse household might not buy any life insurance, even when it is priced actuarially fairly.  We state the following propositions and corollary without proof. 

As in the case of single-premium life insurance, although the optimal consumption and investment strategies depend on the drift and volatility of the risky asset, the optimal death benefit $\bD$ does not.

\prop{4.4} {The optimal amount of life insurance $\bD$ decreases with the premium loading $\bt$ to the extent that for $\bt$ large enough, buying no life insurance is optimal.}

It is no surprise that the optimal amount death benefit decreases as the loading on the continuous premium increases.  Similarly, we expect that the optimal death benefit increases as the household becomes more risk averse, and this is indeed the case, which we state in the next proposition.

\prop{4.5} {The optimal amount of life insurance $\bD$ increases with the risk aversion of the household, as measured by $\al$.}

As in Section 3.2, we have the following corollary to Proposition 4.5.

\cor{4.6} {For $\bt \ge 0$, there exists $\al(\bt) > 0$ such that purchasing no insurance is optimal if and only if $\al \le \al(\bt)$.}

In other words, even for actuarially fair insurance (that is, $\bt = 0$), if the household is not very risk averse, then it will not buy life insurance, a result that we did not expect in this case nor in the case of single-premium life insurance.

In the next proposition, we find that the same conditions as stated in Proposition 3.8 ensure that the optimal amount of life insurance is concave with respect to $\al$.

\prop{4.7} {If $(I_x, \lx) \ge (I_y, \ly)$ or if $(I_x, \lx) \le (I_y, \ly)$, then the optimal amount of life insurance $\bD$ is concave with respect to the risk aversion of the household, as measured by $\al$, when $\bD > 0$.}

Next, we observe that $\bD > 0$ is increasing and convex with respect to the incomes of the household; we state this result without proof because it is clear from the expression for $\bD$ in (4.4).

\prop{4.8} {When $\bD > 0$, the optimal amount of life insurance is increasing and convex with respect to $I_x$ and $I_y$.}

The sign of the derivative of $\bD$ with respect to either force of mortality is ambiguous.  Indeed, from the expression for $\bD$ in (4.4), we see that increasing $\lx$ or $\ly$ shortens the expected horizon, which tends to increase $\bD$ via the logarithmic term; on the other hand, increasing either force of mortality increases the cost of life insurance, which tends to decrease $\bD$ via the $h$-terms.

We end this section with an upper bound on the life insurance purchased by a household.

\prop{4.9} {$$
\bD \le {\max (I_x, I_y) \over h + r} \le \max(I_x, I_y) \, \bar a_{xy}.
$$ That is, $\bD$ is bounded above by the present value of a joint life annuity on $(xy)$ that pays at the rate of $\max(I_x, I_y)$. Moreover, $\lim_{\al \to \infty} \bD = \max(I_x, I_y) \, \bar a_{xy}$.}

Thus, we see that an upper bound for $\bD$ is the actuarial present value of a joint life annuity that pays at the rate of $\max(I_x, I_y)$; this contrasts with the result of Proposition 3.10, namely, $D^* \le \max(I_x, I_y) \, \bar a_{\overline{\infty}|}$.

\sect{5. Actuarial Considerations}

In this section, we consider two problems inspired by actuarial mathematics.  The first is to compare the rates of consumption when the single premium and continuous premium are calculated to ensure a given probability of loss per policy.  The second problem is to find the probability that consumption reaches zero, which relates to the hitting time of Brownian motion with drift, a basic problem in ruin theory.

\subsect{5.1. Targeting a Specific Probability of Loss}

For a given household, we compare the rates of consumption when the single premium $H$ and continuously paid premium $h$ are determined by a given probability of loss.  Assume that the household begins with $W_{0-} = w$ wealth and $D_{0-} = 0$ death benefit and that the optimal amounts of death benefit for both cases are positive, that is, $D^* > 0$ and $\bD > 0$.

In the case of a single premium, the household immediately spends $H D^*$ so that $W_0 = w - HD^*$ and $D_0 = D^*$.  In fact, $D_t = D^*$ for all $t \ge 0$.  By following the wealth dynamics in (2.2) for the optimally controlled wealth, as stated in Theorem 3.3 before the first death and as given in equations (2.6) and (2.7) after the first death, we obtain the optimally controlled wealth for premium payable as a lump sum:
$$
W^*_t =
\cases{w - HD^* + \delta t + {\mu - r \over \al r \sigma} B_t, &$t < \tau_1$. \cr
W^*_{\tau_1} + {m - \la_a \over \al r}(t - \tau_1) + {\mu - r \over \al r \sigma} \left(B_t - B_{\tau_1} \right), &$\tau_1 < t < \tau_2$,}
\eqno(5.1)
$$
in which
$$
\delta = {2 m \over \al r} + I_x  + I_y + {1 \over \al} \ln k(D^*).
\eqno(5.2)
$$
The corresponding optimal rate of consumption is given by
$$
c^*_t =
\cases{r(w - HD^*) + r \delta t + {\mu - r \over \al \sigma} B_t - {1 \over \al} \ln k(D^*), &$t < \tau_1$, \cr
r(w - HD^*) + rD^* + r \delta \tau_1 + {m - \la_a \over \al}(t - \tau_1) + {\mu - r \over \al \sigma} B_t + I_a + {\la_a + m \over \al r}, &$\tau_1 < t < \tau_2$.}
\eqno(5.3)
$$

Similarly, the optimally controlled wealth for premium payable continuously until the first death equals
$$
W^*_t =
\cases{w + \bdel t + {\mu - r \over \al r \sigma} B_t, &$t < \tau_1$. \cr
W^*_{\tau_1} + {m - \la_a \over \al r}(t - \tau_1) + {\mu - r \over \al r \sigma} \left(B_t - B_{\tau_1} \right), &$\tau_1 < t < \tau_2$,}
\eqno(5.4)
$$
in which
$$
\bdel = {2 m \over \al r} + I_x  + I_y + {1 \over \al} \ln \bk(\bD) - h \bD.
\eqno(5.5)
$$
The corresponding optimal rate of consumption is given by
$$
\bc^*_t =
\cases{rw + r \bdel t + {\mu - r \over \al \sigma} B_t - {1 \over \al} \ln \bk(\bD), &$t < \tau_1$, \cr
rw + r \bD + r \bdel \tau_1 + {m - \la_a \over \al}(t - \tau_1) + {\mu - r \over \al \sigma} B_t + I_a + {\la_a + m \over \al r}, &$\tau_1 < t < \tau_2$.}
\eqno(5.6)
$$

Next, we suppose that the single premium $H$ and the premium $h$ are chosen by the insurance company to target a specific probability of loss $q$ on each policy. The loss-at-issue random variables are $L = D^* e^{-r \tau_1} - H D^*$ and $\overline L = \bD e^{-r \tau_1} - h \bD \bar a_{\overline{\tau_1}|}$, respectively.  Because $\tau_1$ is exponentially distributed with mean $1/(\lx + \ly)$, it is straightforward to show that $q = \P(L > 0)$ implies that
$$
q = 1 - H^{{\lx + \ly \over r}},
$$
or equivalently,
$$
H = (1 - q)^{{r \over \lx + \ly}}.
\eqno(5.7)
$$
To ensure that $H \ge {\lx + \ly \over \lx + \ly + r}$, we require that $q \le 1 - \left( {\lx + \ly \over \lx + \ly + r} \right)^{{\lx + \ly \over r}}$.

Similarly, one can show that $q = \P(\overline L > 0)$ implies that
$$
q = 1 - \left( {h \over h + r} \right)^{{\lx + \ly \over r}},
$$
or equivalently,
$$
h = {r (1 - q)^{{r \over \lx + \ly}} \over 1 - (1 - q)^{{r \over \lx + \ly}}}.
\eqno(5.8)
$$
To ensure that $h \ge \lx + \ly$, we require that $q \le 1 - \left( {\lx + \ly \over \lx + \ly + r} \right)^{{\lx + \ly \over r}}$, which is the {\it identical} condition for $H \ge {\lx + \ly \over \lx + \ly + r}$.  Therefore, throughout the rest of this discussion, we assume that
$$
q \le 1 - \left( {\lx + \ly \over \lx + \ly + r} \right)^{{\lx + \ly \over r}}.
\eqno(5.9)
$$

The proof of the next lemma is easy, so we omit it.

\lem{5.1} {If $H$ and $h$ are given by equations $(5.7)$ and $(5.8)$, respectively, then $h = {rH \over 1 - H}$ and $H = {h \over h + r}$.}

The next lemma we be useful in proving our main result concerning $c^*_t$ and $\bc^*_t$.

\lem{5.2} {If $H$ and $h$ are given by equations $(5.7)$ and $(5.8)$, respectively, then $\delta = \bdel$.}

\pf From the right-hand side of inequality (3.4), we have
$$
\ln k(D^*) = {H \over 1- H} - \al (I_x + I_y) - {\lx + \ly + m \over r},
\eqno(5.10)
$$
and similarly,
$$
\ln \bk(\bD) = {h \over r} + \al h \bD - \al (I_x + I_y) -  {\lx + \ly + m \over r}.
\eqno(5.11)
$$
Thus, by substituting (5.10) and (5.11) into (5.2) and (5.5), respectively, we obtain
$$
\delta = {m - \lx - \ly \over \al r} + {1 \over \al} {H \over 1 - H},
\eqno(5.12)
$$
and
$$
\bdel = {m - \lx - \ly \over \al r} + {1 \over \al} {h \over r}.
\eqno(5.13)
$$
It follows from Lemma 5.1 that $\delta = \bdel$.  \qed

We give the main result of this section in the following theorem.

\thm{5.3} {If $H$ and $h$ are given by equations $(5.7)$ and $(5.8)$, respectively, then $c^*_t = \bc^*_t$ for all $0 \le t < \tau_2$ with probability $1$.}

\pf  From equations (5.3) and (5.6), we learn that $c^*_t = \bc^*_t$ for all $t < \tau_1$ if and only if $r H D^* + {1 \over \al} \ln k(D^*) = {1 \over \al} \ln \bk(\bD)$, in which we use the fact that $\delta = \bdel$.  This condition reduces to $r H D^* = h \bD$, which follows from equations (3.5) and (4.4) by using $h = {rH \over 1 - H}$ and $H = {h \over h + r}$ from Lemma 5.1.  Thus, $c^*_t = \bc^*_t$ for all $t < \tau_1$.

We proceed similarly to show $c^*_t = \bc^*_t$ for all $\tau_1 < t < \tau_2$; this equality is equivalent to $(1- H) D^* = \bD$, which follows from equations (3.5) and (4.4) by using Lemma 5.1.  We have, thus, proved the theorem.  \qed

From the proof of the above theorem, we note that $D^* > \bD$; however, because the rates of consumption are identical under either premium-payment scheme, the household will be indifferent between the two.  We have the following corollary to Theorem 5.3.

\cor{5.4} {If $\theta = \bt = 0$, then $c^*_t = \bc^*_t$ for all $0 \le t < \tau_2$ with probability $1$.}

\pf The condition that $\theta = \bt = 0$ is equivalent to $H$ and $h$ determined by targeting the probability of loss equal to the right-hand side of inequality (5.9).  \qed

\rem{5.1} We find it quite surprising that the two rates of consumption are {\it identical} when the single premium and premium rate are computed by a loss-percentile principle. One might be tempted to say that two risk loadings $\theta$ and $\bt$ are equivalent if they result in the same optimal death benefits, namely, $D^* = \bD$.  However, because the household is maximizing its utility of consumption, it makes more sense to say that two risk loadings are equivalent when they result in the same rates of consumption, namely, $c^*_t = \bc^*_t$ for all $t \ge 0$. Theorem 5.3 tells us that we achieve that type of equivalence when the probability of loss for the insurance company is the same under the two premium schemes. So, not only is there equivalence for the buyers of insurance, the {\it insurers} see the two schemes as equivalent from the standpoint of probability of loss.

\subsect{5.2. Probability of Zero Consumption}

Because we do not restrict the rate of consumption to be non-negative, it is of interest to know the probability that consumption reaches zero.  We determine this probability for a household buying life insurance via a single premium or via continuously paid premium.

To calculate the probability that consumption reaches zero, we assume that initially the household has not bought any life insurance, as in Section 5.1, then purchases the optimal amount of life insurance, either $D^*$ or $\bD$, depending on whether the premium is payable at time zero or continuously until the first death, respectively.   Correspondingly define $\tau_0 = \inf \{ t \ge 0: c^*_t \le 0 \}$, in which $c^*_t$ is given in (5.3); similarly, define $\btau_0 = \inf \{ t \ge 0: \bc^*_t \le 0 \}$, in which $\bc^*_t$ is given in (5.6).

First, assume that the premium is payable at time zero, and calculate the probability $\P(\tau_0 < \tau_1 \, | \, c^*_0 > 0)$, in which $c^*_0 = r(w - HD^*) -{1 \over \al} \ln k(D^*)$.  From Corollary B.3.4 in Musiela and Rutkowski (1997), we deduce that the probability density function for $\tau_0 < \infty$ when $c^*_t  = c^*_0 + r \delta t +  {\mu - r \over \al \sigma} B_t$ equals
$$
f_{\tau_0} (t) = {c^*_0 \over {\mu - r \over \al \sigma} \sqrt{2 \pi t^3}} \exp \left[ - {\al^2 \over m t} \, (c^*_0 + r \delta t)^2 \right],
\eqno(5.14)
$$
with $\int_0^\infty f_{\tau_0} (t) \, dt = e^{-\al^2 r \delta c^*_0/m} < 1$ because $\tau_0$ has a positive probability of taking the value infinity.  Recall that $m = {1 \over 2} \left( {\mu - r \over \sigma} \right)^2$, so $ \left( {\mu - r \over \al \sigma} \right)^2 = {2 m \over \al^2}$.  It follows that
$$
\P(\tau_0 < \tau_1 \, | \, c^*_0 > 0) = \int_0^\infty f_{\tau_0} (t) \, e^{-(\lx + \ly)t} \, dt = \exp \left[ -{\al^2 c^*_0 \over 2 m} \left(  \sqrt{(r \delta)^2 + {4m \over \al^2} \, (\lx + \ly)} - r \delta \right) \right].
\eqno(5.15)
$$

The probability $\P(\tau_1 < \tau_0 < \tau_2 \, | \, c^*_0 > 0)$ is rather complicated because it depends on whether the change in the rate of consumption, $\Delta c^* = \Delta c(D^*)$ from Corollary 3.4, is positive and if it is positive, whether it is greater than $c^*_0$. Also, note that $\Delta c^*$ depends on whether $(x)$ or $(y)$ dies first, so we write $\Delta c^*_a$ with $a = x$ if $(y)$ dies first and with $a = y$ if $(x)$ dies first.  For the sake of brevity, we include only an outline of how one calculates $\P(\tau_1 < \tau_0 < \tau_2 \, | \, c^*_0 > 0)$; then, we write its expression in the Appendix for the interested reader.

If we are given $\tau_1 =  t$ and $\tau_0 > t$, then the expression in (5.15), with the drift $r \delta$ replaced by ${m - \la_a \over \al}$ and the force of mortality $\lx + \ly$ replaced by $\la_a$, implies that
$$
\P(\tau_0 < \tau_2 \, | \, \tau_1 = t, \tau_0 > t, c^*_t) = e^{-\al c^*_t}.
\eqno(5.16)
$$
From Corollary B.3.4 in Musiela and Rutkowski (1997), we deduce that the joint probability density function of $c^*_{t-}$ and $\min_{0 \le u < t} c^*_u$ equals
$$
g(c, n) = {2(c - 2n + c^*_0) \over \left( {\mu - r \over \al \sigma} \right)^3 \, t^{3/2}} \exp \left( {\al^2 r \delta (n - c^*_0) \over m} \right) \, \phi \left( {c - 2n + c^*_0 - r \delta t \over {\mu - r \over \al \sigma} \sqrt{t}} \right), 
$$
in which $\phi$ is the probability density function of the standard normal random variable.  The domain of $g$ is $\{ (c, n): c \ge n, n \le c^*_0 \}$.  Note that if $\tau_1 = t$, then $c^*_t = c^*_{t-} + \Delta c^*_a$, which might be greater than or less than $c^*_{t-}$ depending on the sign of $\Delta c^*_a$.  It follows that
$$
\eqalign{
\P(\tau_1 < \tau_0 < \tau_2 \, | \, c^*_0 > 0) &= \int_0^\infty \int_{\cal N} \int_{\cal C} \, g(c, n) \, e^{-\al (c + \Delta c^*_x)} \, \ly \, e^{-(\lx + \ly) t} \, dc \, dn \, dt \cr
& \quad + \int_0^\infty \int_{\cal N} \int_{\cal C} \, g(c, n) \, e^{-\al (c + \Delta c^*_y)} \, \lx \, e^{-(\lx + \ly) t} \, dc \, dn \, dt,}
\eqno(5.17)
$$
in which we integrate over the two regions in the $c$-$n$ plane: $\{ (c, n): c > (- \Delta c^*_a)_+, c \ge n, 0 \le n \le c^*_0 \}$, the first integral with $a = x$ and the second with $a = y$. Please see equations (A.1)-(A.6) in the Appendix for explicit expressions of this probability.

To compute the probability that consumption reaches zero before the second death add the probabilities $\P(\tau_0 < \tau_1 \, | \, c^*_0 > 0)$ and $\P(\tau_1 < \tau_0 < \tau_2 \, | \, c^*_0 > 0)$.  Without loss of generality, suppose that $\Delta c^*_y \le \Delta c^*_x$.  As a function of $c^*_0$, we have three cases to consider in computing $\P(\tau_0 < \tau_2 \, | \, c^*_0 > 0)$, depending on the relationship of $-\Delta c^*_x \le -\Delta c^*_y$ with 0.  Please see equations (A.7)-(A-9) in the Appendix for these three expressions.

Now, assume that the premium is payable continuously.  By analogy with equation (5.15), the probability that consumption reaches zero before the first death equals
$$
\P(\btau_0 < \tau_1 \, | \, \bc^*_0 > 0) = \exp \left[ -{\al^2 \, \bc^*_0 \over 2 m} \left( \sqrt{(r \bdel)^2 + {4m \over \al^2} \, (\lx + \ly)} - r \bdel \right) \right],
\eqno(5.18)
$$
in which $\bc^*_0 = rw - {1 \over \al} \ln \bk(\bD)$.  For the probability that $\btau_0$ occurs after the first death and before the second, we have the same six cases as detailed in the Appendix.  For brevity, we do not write down that probability in those cases because the expressions are analogous to those in equations (A.1)-(A.9) with $c^*_0$, $\delta$, and $\Delta c^*_a$ replaced by $\bc^*_0$, $\bdel$, and $\Delta \bc^*_a = \Delta \bc(\bD)$, respectively.

\sect{6. Numerical Example}

We present a numerical example to illustrate some of the results from the previous sections.  First, set the parameters for the financial and insurance market in this base scenario:

\smallskip

\item{$\bullet$} The riskless rate of return, or force of interest, is $r = 0.02$.

\item{$\bullet$} The drift of the risky asset is $\mu = 0.06$.

\item{$\bullet$} The volatility of the risky asset is $\sigma = 0.20$.

\item{$\bullet$} The proportional risk loadings are both zero:  $\theta = \bt = 0$.

\smallskip

\noi Thus, $m = {1 \over 2} \left( {\mu - r \over \sigma} \right)^2 = 0.02$.  Next, set the parameters for the household:

\smallskip

\item{$\bullet$} The force of mortality of $(x)$ is $\lx = 0.04$, so $\E \tau_x = 25$.

\item{$\bullet$} The force of mortality of $(y)$ is $\ly = 0.03$, so $\E \tau_y = 33 {1 \over 3}$.

\item{$\bullet$} The income rate of $(x)$ is $I_x = 2.0$, in multiples of \$50,000, so the expected lifetime income is $2(25) = 50$ or \$2,500,000.

\item{$\bullet$} The income rate of $(y)$ is $I_y = 1.5$, in multiples of \$50,000, so the expected lifetime income is also $1.5(100/3) = 50$ or \$2,500,000.

\item{$\bullet$} The absolute risk aversion of the household is $\al = 2$, which means that the household would pay \$122.65 to insure against a loss of \$10,000 (0.2 of \$50,000) with probability 0.01.  The expected loss is \$100, so willingness to pay \$122.65 to protect against this random loss seems reasonable.

\smallskip
  
\noi It follows that, per dollar of life insurance, the single premium is $H = {7 \over 9}$, and the continuous premium rate is $h = 0.07$. From (3.5), we compute the optimal death benefit in the case of single premium life insurance: $D^* = 52.38$ or \$2,619,000.  Similarly, from (4.4), we compute the optimal death benefit when premium is payable continuously: $\bD = 11.64$ or \$582,000.  Despite the large difference in death benefit amounts, we know that these benefits are ``equivalent'' in the sense of providing the household with the same rates of consumption; recall Corollary 5.4. The corresponding probability of loss per policy for the insurance company is $q = 1 - (7/9)^{3.5} = 0.585$.

The change in the rate of consumption at the first death equals $\Delta c^*_x = 0.5476$ or \$27,380 if $(y)$ dies first and $\Delta c^*_y = -0.2024$ or $-$\$10,120 if $(x)$ dies first.  It is not a surprise that the rate of consumption of the household decreases when $(x)$ dies first because of the loss of $(x)$'s greater income, although the death benefit $D^* = \;$\$2,619,000 ameliorates the decrease.  The changes in the rate of consumption are identical for the case of premium payable continuously, as shown in Corollary 5.4.



By using the expressions developed in Section 5.2, we conclude from several numerical experiments that the probability that consumption reaches zero is so small that one can effectively ignore the anomaly of consumption becoming negative.  In other words, the cost of the mathematical work involved in imposing the constraint that consumption rates be non-negative is not worth the benefit of determining the corresponding optimal strategies because they will differ very little from those that we compute in this paper.  

\sect{7. Summary and Conclusions}

We determined the optimal amount of life insurance for a household of two wage earners.  We considered the simple case of exponential utility, thereby removing wealth as a factor in buying life insurance, while retaining the relationship among life insurance, income, and the probability of dying and thereby losing that income. For insurance purchased via a single premium or premium payable continuously, we explicitly determined the optimal death benefit; see equations (3.5) and (4.4), respectively.

For the reader who wishes to have a handy reference without having to search through the paper, we reproduce those equations here, along with a list of what the notation means.  The optimal amount of death benefit, if insurance is charge via a single premium $H$ per dollar of insurance, equals
$$
D^* = \max \left[ {1 \over \al r} \left[ \ln \left( \lx \exp \left(\al I_x + {\lx \over r} \right) + \ly \exp \left(\al I_y + {\ly \over r} \right) \right) - \ln \left( {rH \over 1- H} \right) - {H \over 1 - H} \right], \, 0 \, \right].
$$
Here, $\al$ is the (constant) absolute risk aversion of the household, $\lx$ and $\ly$ are the forces of mortality for $(x)$ and $(y)$, $I_x$ and $I_y$ are the rates of income of $(x)$ and $(y)$, and $r$ is the force of interest.  The optimal amount of death benefit, if premium is payable continuously until the first death at a rate of $h$ per dollar of insurance, equals
$$
\bD = \max \left[ {1 \over \al (h + r)} \left[ \ln \left( \lx \exp \left(\al I_x + {\lx \over r} \right) + \ly \exp \left(\al I_y + {\ly \over r} \right) \right) - \ln h - {h \over r} \right], \, 0 \, \right].
$$
See the Excel spreadsheet at

http://www.math.lsa.umich.edu/$\sim$erhan/OptimalLifeInsurance.xlsx

\noi for a tool that one can use to calculate either $D^*$ or $\bD$.

We determined how the optimal death benefit varies with the underlying parameters, and we showed that the death benefits are bounded above by annuities based on the maximum income lost; see Propositions 3.10 and 4.9. From those propositions, we concluded that the household does not over-insure, a desirable outcome of any model.

In Section 5, we showed that if the premium is determined to target a specific probability of loss per policy, then the rates of consumption are identical.  Thus, not only is equivalence of consumption achieved for the households under the two premium schemes, it is also obtained for the insurance company in the sense of equivalence of loss probabilities.

We welcome the interested reader to modify our model for other common configurations of households, as described in Remarks 2.1 and 2.2.  In future work, we plan to explore optimal life insurance purchasing for someone who wishes to bequeath a certain amount of money to a beneficiary.

\bigskip

\centerline{\bf Acknowledgments} \medskip  We thank the Committee for Knowledge Extension and Research of the Society of Actuaries for financially supporting this work.  Additionally, research of the first author is supported in part by the National Science Foundation under grants DMS-0955463, DMS-0906257, and the Susan M. Smith Professorship of Actuarial Mathematics. Research of the second author is supported in part by the Cecil J. and Ethel M. Nesbitt Professorship of Actuarial Mathematics.

\sect{Appendix}

Without loss of generality, suppose that $\Delta c^*_y \le \Delta c^*_x$, or equivalently, $\al I_y + {\ly \over r} \le \al I_x + {\lx \over r}$.    We have six cases to consider, and we provide $P(c^*_0) = \P(\tau_1 < \tau_0 < \tau_2 \, | \, c^*_0 > 0)$ for each case.  In fact, we only have five cases to consider because one can show that $\Delta c^*_y \le 0$ automatically.  However, we include case A below for completeness.  Technically, case A includes the possible case of $\Delta c^*_y = 0$, but one could compute the corresponding probability by letting $\Delta c^*_y \to 0$ in equation (A.2). 

To simplify the expressions, define
$$
S = \sqrt{(r \delta)^2 + {4 m \over \al^2} \, (\lx + \ly) }.
$$

\item{A. } $-\Delta c^*_x \le -\Delta c^*_y \le 0 < c^*_0$.
$$
P_A(c^*_0) = {\ly e^{-\al \Delta c^*_x} + \lx e^{-\al \Delta c^*_y} \over \left( {\al \over 2m} (S - r \delta) + 1 \right) \left( {1 \over 2} \al r \delta - m \right)} e^{- {\al^2 \over 2m}(S + r \delta) c^*_0} \left( e^{ \left( {\al^2 r \delta \over 2 m} - \al \right) c^*_0} - 1 \right).
\eqno(A.1)
$$

\item{B. } $-\Delta c^*_x \le 0 < -\Delta c^*_y < c^*_0$.
$$
\eqalign{
P_B(c^*_0) &= {\ly e^{-\al \Delta c^*_x} \over \left( {\al \over 2m} (S - r \delta) + 1 \right) \left( {1 \over 2} \al r \delta - m \right)} e^{- {\al^2 \over 2m}(S + r \delta) c^*_0} \left( e^{ \left( {\al^2 r \delta \over 2 m} - \al \right) c^*_0} - 1 \right) \cr
& \quad + {\lx e^{- {\al^2 \over 2m} (S + r \delta) (c^*_0 + \Delta c^*_y)} \over {\al^2 \over 2m} (S - r \delta) + \al}  \left[ {1 \over S} \left( 1 - e^{{\al^2 S \over m} \Delta c^*_y} \right) + {\al \, e^{{\al^2 S \over 2 m} \Delta c^*_y} \over {1 \over 2} \al r \delta - m}  \left( e^{ \left( {\al^2 r \delta \over 2m} - \al \right) (c^*_0 + \Delta c^*_y)} - 1 \right) \right].}
\eqno(A.2)
$$

\item{C. } $0 < -\Delta c^*_x \le -\Delta c^*_y < c^*_0$.
$$
\eqalign{
P_C(c^*_0) &=  {\ly e^{- {\al^2 \over 2m} (S + r \delta) (c^*_0 + \Delta c^*_x)} \over {\al^2 \over 2m} (S - r \delta) + \al}  \left[ {1 \over S} \left( 1 - e^{{\al^2 S \over m} \Delta c^*_x} \right) + {\al \, e^{{\al^2 S \over 2 m} \Delta c^*_x} \over {1 \over 2} \al r \delta - m}  \left( e^{ \left( {\al^2 r \delta \over 2m} - \al \right) (c^*_0 + \Delta c^*_x)} - 1 \right) \right] \cr
& \quad + {\lx e^{- {\al^2 \over 2m} (S + r \delta) (c^*_0 + \Delta c^*_y)} \over {\al^2 \over 2m} (S - r \delta) + \al}  \left[ {1 \over S} \left( 1 - e^{{\al^2 S \over m} \Delta c^*_y} \right) + {\al \, e^{{\al^2 S \over 2 m} \Delta c^*_y} \over {1 \over 2} \al r \delta - m}  \left( e^{ \left( {\al^2 r \delta \over 2m} - \al \right) (c^*_0 + \Delta c^*_y)} - 1 \right) \right].}
\eqno(A.3)
$$

\item{D. } $-\Delta c^*_x \le 0 < c^*_0 \le -\Delta c^*_y$.
$$
\eqalign{
P_D(c^*_0) &= {\ly e^{-\al \Delta c^*_x}  \over \left( {\al \over 2m} (S - r \delta) + 1 \right) \left( {1 \over 2} \al r \delta - m \right)} e^{- {\al^2 \over 2m}(S + r \delta) c^*_0} \left( e^{ \left( {\al^2 r \delta \over 2 m} - \al \right) c^*_0} - 1 \right) \cr
& \quad + {\lx \over S \left(  {\al^2 \over 2m} (S - r \delta) + \al \right)} e^{ {\al^2 \over 2m} (S - r \delta) (c^*_0 + \Delta c^*_y)} \left( 1 - e^{-{\al^2 S \over m} c^*_0} \right).}
\eqno(A.4)
$$

\item{E. } $0 < -\Delta c^*_x  < c^*_0 \le -\Delta c^*_y$.
$$
\eqalign{
P_E(c^*_0) &= {\ly e^{- {\al^2 \over 2m} (S + r \delta) (c^*_0 + \Delta c^*_x)} \over {\al^2 \over 2m} (S - r \delta) + \al}  \left[ {1 \over S} \left( 1 - e^{{\al^2 S \over m} \Delta c^*_x} \right) + {\al \, e^{{\al^2 S \over 2 m} \Delta c^*_x} \over {1 \over 2} \al r \delta - m}  \left( e^{ \left( {\al^2 r \delta \over 2m} - \al \right) (c^*_0 + \Delta c^*_x)} - 1 \right) \right] \cr
& \quad + {\lx \over S \left(  {\al^2 \over 2m} (S - r \delta) + \al \right)} e^{ {\al^2 \over 2m} (S - r \delta) (c^*_0 + \Delta c^*_y)} \left( 1 - e^{-{\al^2 S \over m} c^*_0} \right).}
\eqno(A.5)
$$

\item{F. } $0 < c^*_0 \le -\Delta c^*_x \le -\Delta c^*_y$.
$$
P_F(c^*_0) = {1 - e^{-{\al^2 S \over m} c^*_0}  \over S \left(   {\al^2 \over 2m} (S - r \delta) + \al \right)} \left[ \ly e^{{\al^2 \over 2m} ( S - r \delta) (c^*_0 + \Delta c^*_x)} + \lx e^{{\al^2 \over 2m} (S - r \delta) (c^*_0 + \Delta c^*_y)} \right].
\eqno(A.6)
$$

\medskip

Next, we compute $p(c^*_0) = \P(\tau_0 < \tau_2 \, | \, c^*_0 > 0)$ for the three cases corresponding to the possible relationships of $-\Delta c^*_x \le -\Delta c^*_y$ with 0.

\smallskip

\item{I. } $-\Delta c^*_x \le -\Delta c^*_y \le 0$.
$$
p(c^*_0) = e^{-{\al^2 \over 2m}(S - r \delta) c^*_0}+P_A(c^*_0),
\eqno(A.7)
$$
in which $P_A$ is given in equation (A.1).

\item{II. } $-\Delta c^*_x \le 0 < -\Delta c^*_y$.
$$
p(c^*_0) = 
\cases{ e^{-{\al^2 \over 2m}(S - r \delta) c^*_0}+P_D(c^*_0), &if $0 < c^*_0 \le -\Delta c^*_y$, \cr
e^{-{\al^2 \over 2m}(S - r \delta) c^*_0}+P_B(c^*_0), &if $c^*_0 > -\Delta c^*_y$.}
\eqno(A.8)
$$
in which $P_D$ and $P_B$ are given in equations (A.4) and (A.2), respectively.

\item{III. }  $0 < -\Delta c^*_x \le -\Delta c^*_y$.
$$
p(c^*_0) = 
\cases{ e^{-{\al^2 \over 2m}(S - r \delta) c^*_0}+P_F(c^*_0), &if $0 < c^*_0 \le -\Delta c^*_x$, \cr
e^{-{\al^2 \over 2m}(S - r \delta) c^*_0}+P_E(c^*_0), &if $-\Delta c^*_x < c^*_0 \le -\Delta c^*_y$, \cr
e^{-{\al^2 \over 2m}(S - r \delta) c^*_0}+P_C(c^*_0), &if $c^*_0 > -\Delta c^*_y$.}
\eqno(A.9)
$$
in which $P_F$, $P_E$, and $P_c$ are given in equations (A.6), (A.5), and (A.3), respectively.

\sect{References}


\noindent \hangindent 20 pt Bowers, Newton L., Hans U. Gerber, James C. Hickman, Donald A. Jones, and Cecil J. Nesbitt (1997), {\it Actuarial Mathematics}, second edition, Schaumburg, IL: Society of Actuaries.

\smallskip \noindent \hangindent 20 pt Bruhn, Kenneth and Mogens Steffensen (2011), Household consumption, investment and life insurance, {\it Insurance: Mathematics and Economics}, 48 (3): 315-325.

\smallskip \noindent \hangindent 20 pt Campbell, Ritchie A. (1980), The demand for life insurance: an application of the economics of uncertainty, {\it Journal of Finance}, 35 (5), 1155-1172.


\smallskip \noindent \hangindent 20 pt Egami, Masahiko and Hideki Iwaki (2011), An optimal life insurance purchase in the investment-consumption problem in an incomplete market, working paper, arXiv:0801.0195v2.

\smallskip \noindent \hangindent 20 pt Frees, Edward W., Jacques Carriere, and Emiliano A. Valdez (1996), Annuity valuation with dependent mortality, {\it Journal of Risk and Insurance}, 63 (2): 229-261.

\smallskip \noindent \hangindent 20 pt Gerber, Hans U. (1974), On additive premium calculation principles, {\it ASTIN Bulletin}, 7 (3): 215-222.

\smallskip \noindent \hangindent 20 pt Huang, Huaxiong and Moshe A. Milevsky (2008), Portfolio choice and mortality-contingent claims: the general HARA case, {\it Journal of Banking and Finance}, 32 (11): 2444-2452.


\smallskip \noindent \hangindent 20 pt  Jeanblanc, Monique, Peter Lakner, and Kadam (2004), Optimal bankruptcy time and consumption/ \break investment policies on an infinite horizon with a continuous debt repayment until bankruptcy, {\it Mathematics of Operations Research}, 29 (3): 649-671.


\smallskip \noindent \hangindent 20 pt Kraft, Holger and Mogens Steffensen (2008), Optimal consumption and insurance: a continuous-time Markov chain approach, {\it ASTIN Bulletin}, 28: 231-257.

\smallskip \noindent \hangindent 20 pt Kwak, Minsuk, Yong Hyun Shin, and U. Jin Choi (2011), Optimal investment and consumption decision of a family with life insurance, {\it Insurance: Mathematics and Economics}, 48 (2): 176-188.

\smallskip \noindent \hangindent 20 pt Merton, Robert C. (1969), Lifetime portfolio selection under uncertainty: the continuous-time case, {\it Review of Economics and Statistics}, 51 (3): 247-257.

\smallskip \noindent \hangindent 20 pt Milevsky, Moshe A. and Virginia R. Young (2007), Annuitization and asset allocation, {\it Journal of Economic Dynamics and Control}, 31 (9): 3138-3177.

\smallskip \noindent \hangindent 20 pt Musiela, Marek and Marek Rutkowski (1997), {\it Martingale Methods in Financial Modelling}, Springer, New York.

\smallskip \noindent \hangindent 20 pt Nielsen, Peter Holm and Mogens Steffensen (2008), Optimal investment and life insurance strategies under minimum and maximum constraints, {\it Insurance: Mathematics and Economics}, 43 (1): 15-28.


\smallskip \noindent \hangindent 20 pt Pliska, Stanley and Jinchun Ye (2007), Optimal life insurance purchase and consumption/investment under uncertain lifetime, {\it Journal of Banking and Finance}, 31 (5): 1307-1319.

\smallskip \noindent \hangindent 20 pt Richard, Scott F. (1975), Optimal consumption, portfolio and life insurance rules for an uncertain lived individual in a continuous time model, {\it Journal of Financial Economics}, 2 (2): 187-203.

\smallskip \noindent \hangindent 20 pt Wang, Jennifer L., H. C. Huang, Sharon S. Yang, and Jeffrey T. Tsai (2010), An optimal product mix for hedging longevity risk in life insurance companies: the immunization theory approach, {\it Journal of Risk and Insurance}, 77 (2): 473-497.

\smallskip \noindent \hangindent 20 pt Wang, Ting and Virginia R. Young (2012a), Optimal commutable annuities to minimize the probability of lifetime ruin, {\it Insurance: Mathematics and Economics}, 50 (1): 200-216.

\smallskip \noindent \hangindent 20 pt Wang, Ting and Virginia R. Young (2012b), Maximizing the utility of consumption with commutable annuities, {\it Insurance: Mathematics and Economics}, 51 (2): 352-369.


 \bye